\documentclass[prl,reprint,twocolumn,showpacs,preprintnumbers,amsmath,amssymb,floatfix]{revtex4}
\usepackage[german,english]{babel}
\usepackage{bm,amsmath,amsfonts}
\usepackage{amssymb,siunitx}
\usepackage{graphicx}
\usepackage{xcolor}

\def\beq{\begin{equation}}
\def\eeq{\end{equation}}
\def\beqn{\begin{eqnarray}}
\def\eeqn{\end{eqnarray}}
\def\souligne#1{$\underline{\smash{ \hbox{#1}}}$}
\def\brho {\mbox{\boldmath $\rho$}}
\def\r {\vec{\bf r}}
\def\r {{\bf r}}
\def\br {{\bf r}}
\def\bR {{\bf R}}
\def\bcalH {\mbox{\boldmath $\mathcal H$}}
\def\bldrho  {\boldsymbol{\rho}}
\def\ul {\underline}

\begin{document}

\title{Generic regimes of quantum many-body dynamics \\ of trapped bosonic systems with strong repulsive interactions}

\author{Oksana I. Streltsova$^1$, Ofir E. Alon$^2$, Lorenz S. Cederbaum$^3$,  
and Alexej I. Streltsov$^3$}

\affiliation{$^1$
Laboratory of Information Technologies,
Joint Institute for Nuclear Research,
Joliot-Curie 6, Dubna, Russia}

\affiliation{$^2$ Department of Physics, University of Haifa at Oranim,
Tivon 36006, Israel}

\affiliation{$^3$Theoretische Chemie, Physikalisch-Chemisches Institut, Universit\"at Heidelberg,\\
Im Neuenheimer Feld 229, D-69120 Heidelberg, Germany}

\begin{abstract}
Two generically different but universal dynamical quantum many-body behaviors are discovered
by probing the stability of trapped fragmented bosonic systems with strong repulsive finite/long range inter-particle interactions.
We use different time-dependent processes to destabilize the systems --
a sudden displacement of the trap is accompanied by a sudden quench of the strength of the inter-particle repulsion.
A rather moderate non-violent evolution of the density in the first ``topology-preserved'' scenario is contrasted  with a highly-non-equilibrium  dynamics characterizing an explosive changes of the density profiles in the second scenario. The many-body physics behind is identified and interpreted in terms of self-induced time-dependent barriers governing the respective under- and over-a-barrier dynamical evolutions. The universality of the discovered scenarios is explicitly confirmed 
in 1D, 2D and 3D  many-body computations 
in (a)symmetric traps and repulsive finite/long range inter-particle interaction potentials of different shapes. Implications are briefly discussed. 
\end{abstract}
\maketitle

One of the most bright universal features shared by many-body systems 
with strong repulsive interaction is the formation of multi-hump structures in the
ground states' densities. Driven by the strong repulsive interaction they can be formed in the systems with short- and finite/long-range inter-particle interactions in one- (1D), two- (2D) and three-dimensional (3D) setups.
The famous examples in the context of ultra-cold systems 
are strong contact interaction and Tonks-Girardeu gases in 1D \cite{Girardeau_TG,Fermionization_Exp} and
-- in more general
physical contents -- the formation of super-solids and crystals in 2D 
systems with long-range interactions \cite{Dip_review0,Dip_review}.
While ground state properties of these systems have been accessed at different levels of the quantum theory, the many-body studies on excited states, needed to digest dynamical behavior and stability of these systems
are rather scarce. Because of the intrinsic complexity and correlations of these systems, the understanding of their dynamical stability as a time-dependent process at a proper many-body level is a challenging theoretical and cumbersome computational task and is still missed.

In this Letter we investigate the stability of such strongly repulsive 
many-boson systems with muti-hump many-body states 
as a time-dependent process by solving the time-dependent many-boson Schr\"odinger equation 
$\hat H \Psi\!=\!i\hbar \frac{\partial\Psi}{\partial t}$ in 1D, 2D and 3D setups
for several dynamical scenarios involving manipulations with external trap $V(\r,t)$ and with the strength $\lambda_0$ of inter-boson interaction 
potential $W(\br\!-\!\br') \equiv W(\bR)$. The respective many-body Hamiltonian is
$\hat H  \!=\! \sum_{j=1}^N \left[- \frac{1}{2} \nabla^2_{\r_j} \!+\! V(\r_j,t) \right]
\!+\!\sum_{j < k}^N \lambda_0 W(\r_j\!-\!\r_k), \nonumber$ $\hbar\!=\!1$, $m\!=\!1$.
All the results reported in this Letter have been obtained 
for $N=100$ bosons interacting via inter-particle interaction function 
$W(\bR)\!=\!1/{((|\br-\br'|/D)^{n}\!+\!1})$ of half-width $D\!=\!4$ with $n\!=\!4$
in 1D, 2D and 3D. The inter-particle interactions of similar shapes  
naturally appear in the so-called ``Rydberg-dressed'' ultra-cold systems \cite{Ry_review,Ry1,Ry2} which are of current experimental interest \cite{Ry3_exp}. The repulsive inter-particle interaction functions of other shapes with similar range and strength would result in qualitatively the same physics as reported here, see also the supplemental material \cite{SM} for discussion.

Recently, we have verified at an accurate many-body level \cite{Towards}
that in bosonic systems with strong repulsion confined in simple, 
barrier-less  traps particular patterns of the ground state density, 
i.e., number of the humps and correlations,
are governed by the geometrical interplay between the width/range of the inter-boson interaction potential and the available volume (length of the trap).
Examples of such two-hump states in 1D and 3D setups are depicted in Fig.~\ref{fig1} and Fig.~\ref{fig3} at t=0.
Intuitively, one can associate each hump of a multi-hump state  with a localized subsystem and to describe, therefore, the overall wave-function
$\Psi(\br_1,\ldots,\br_N,t)$ as a superposition of all constituting fragments.
A configuration where $n_1$ bosons are residing in the first fragment $\phi_1(\r,t)$, 
$n_2-$ in $\phi_2(\r,t)$, etc., and $n_M-$ in $\phi_M(\r,t)$ is called permanent
$|n_1,n_2...,n_M,t\rangle$:
$\Psi(\br_1,\ldots,\br_N,t)\!=\!\hat{\cal  S}
\phi_1(\br_1,\!t)\!\cdots\!\phi_1(\br_{n_1},\!t)
\phi_2(\br_{n_1+1},t)\cdots\phi_M(\br_{N},\!t)$.
Here we have already admitted that the shapes of the constituting sub-clouds can evolve in time.
For a realistic description this idealized picture of a single configuration  
with a fixed number of particle residing in each fragment should
be augmented by other processes describing, e.g., the exchange and hopping of particles between the fragments.

The many-body theory naturally taking into account this time-dependency of the fragments as well as all possible hopping processes within
has recently been developed and called multi-configurational time-dependent Hartree method for Bosons  (MCTDHB)  \cite{ramp_up,MCTDHB_paper}, also see 
Refs.~\cite{fragmneton,BJJ-PRL,swift,MCTDHB_HIM,PNAS} for a few applications.
The MCTDHB many-body wavefunction is a linear combination of all above described {\it time-dependent} permanents
$\Psi(\r_1,\ldots,\r_N,t) = \sum_{\vec{n}} C_{\vec{n}}(t) |n_1,n_2...,n_M,t\rangle.$
The evolutions of the coefficients $C_{\vec{n}}(t)$ describing all possible hopping processes in the system
and the dynamical changes of the shapes of every fragment $\left\{\phi_j(\r,t)\right\}$ are determined by solving the MCTDHB equations:
\begin{eqnarray}\label{MCTDHB_eq}
i{\frac{d C_{\vec{n}}}{dt}}&\!=\!&\sum_{\vec{n}'} \left< \Phi_{\vec{n}}\left|\hat H\right|\Phi_{\vec{n}'}\right> C_{\vec{n}'}(t) \\
i\frac{\partial\phi_j}{\partial t}\!&\!=\!&\!
\hat{\mathbf P}\!\left[\!\hat h
 \phi_j(\r,t)\!+\!\!\sum_{qksl} 
 {\{\bldrho(t)\}^{-1}_{jq}}\rho_{qksl}(t)
 \lambda_0\!W_{kl}(\r,t)\!\phi_s(\r,t)\!\right].\nonumber
\end{eqnarray}
Here $\hat h\!=\!\hat T(\r)\!+\!V(\r,t) $ is the single particle Hamiltonian,
$\bldrho(t)\!=\!\{\rho_{qs}(t)\}$ and $\rho_{qksl}(t)$ are the matrix elements of the reduced one- and two-body densities of $\Psi$.
The local time-dependent potentials $W_{kl}(\r,t)=\int \phi_k^\ast(\r',t)W(\br-\br')\phi_l(\r',t)d\r'$
originate from the two-body interaction and play a crucial role in the physics studied here.
To investigate the dynamical stability of a desired state as a time-dependent process we, first, specify this state by providing corresponding initial conditions $C_{\vec{n}}(t=0)$ and $\left\{\phi_j(\r,t=0)\right\}$,
and then monitor how these quantities change in time in a response 
to the applied modification of the trap $V(\br,t)$  and/or to a quench of the interaction strength  $\lambda_0$ of  $W(\br-\br')$.

Our goal is to investigate the dynamical stability of the multi-hump multi-fold  fragmented states.
In Ref.~\cite{Towards} we have shown that in simple, barrier-less traps
the number of humps and fragmentation ratio of the bosonic system with finite- and long-range repulsive interactions in the ground state can be equivalently controlled by varying either the strength of the inter-particle repulsion $\lambda_0$,
the tightness/strength of harmonic confinement $\omega$,  or by changing the number $N$ of trapped particles.
In the supplemental material \cite{SM}, a detail control of the humps' structures of the ground state in parabolic trap and inter-particle interaction $W(\bR)=1/{((|\br-\br'|/D)^{n}+1})$ of half-width $D=4$ with $n=4$, studied throughout this work, is presented.
So, from now on we assume that a ground state with a desired humps' structures and fragmentation ratio is available.

\begin{figure}[h]
\includegraphics[width=2.5cm,angle=0]{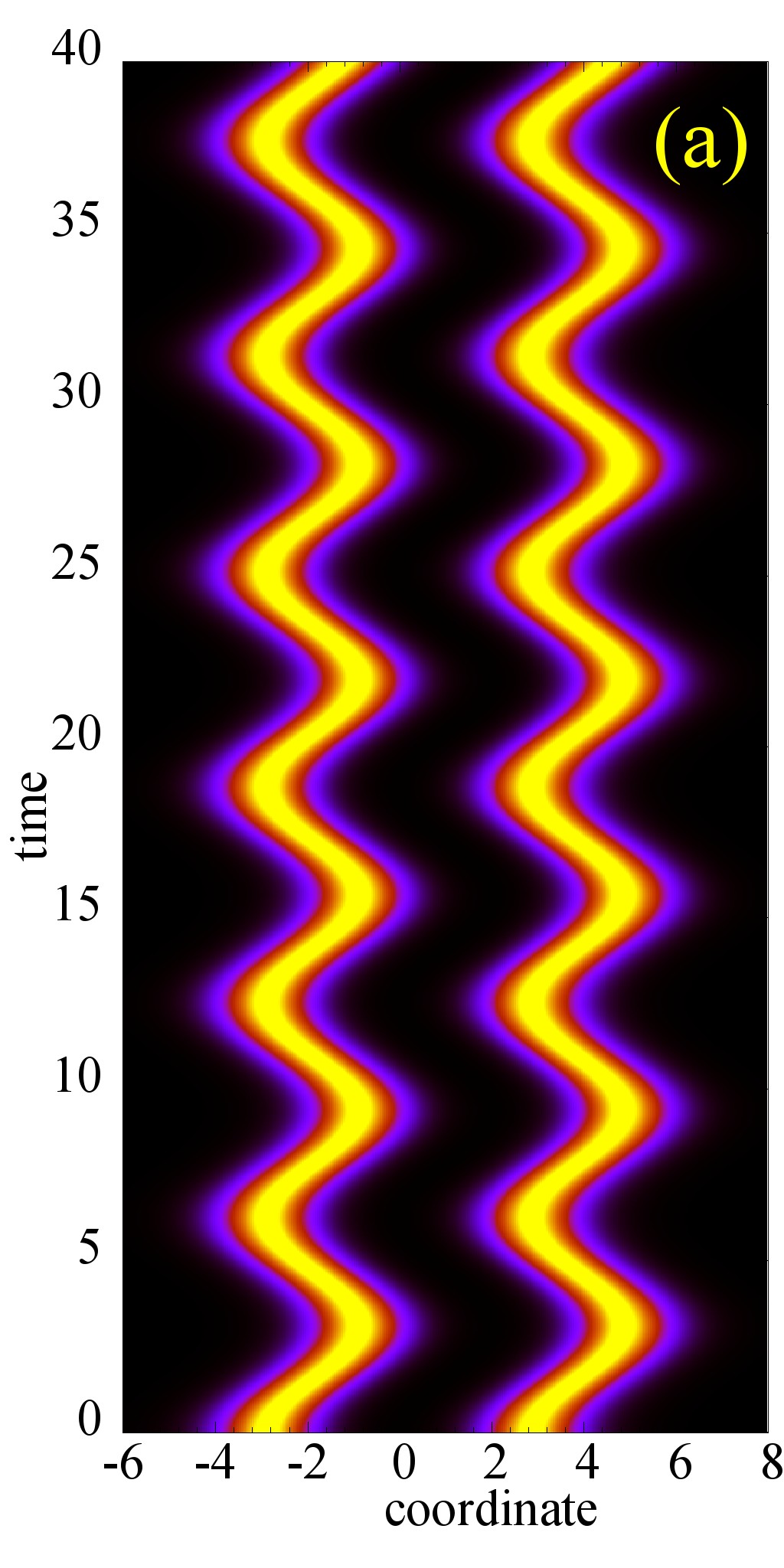}
\includegraphics[width=2.5cm,angle=0]{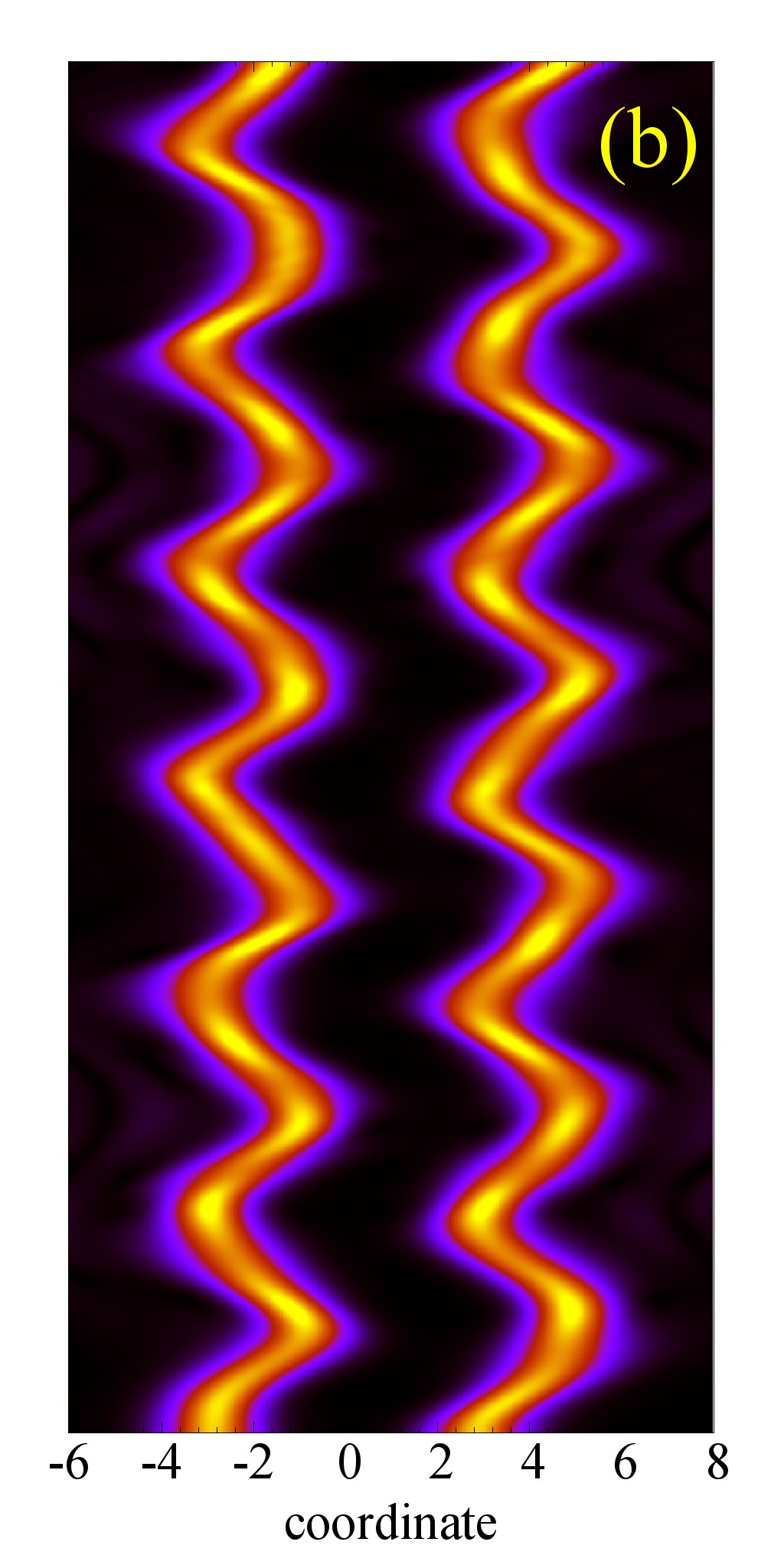}
\includegraphics[width=2.5cm,angle=0]{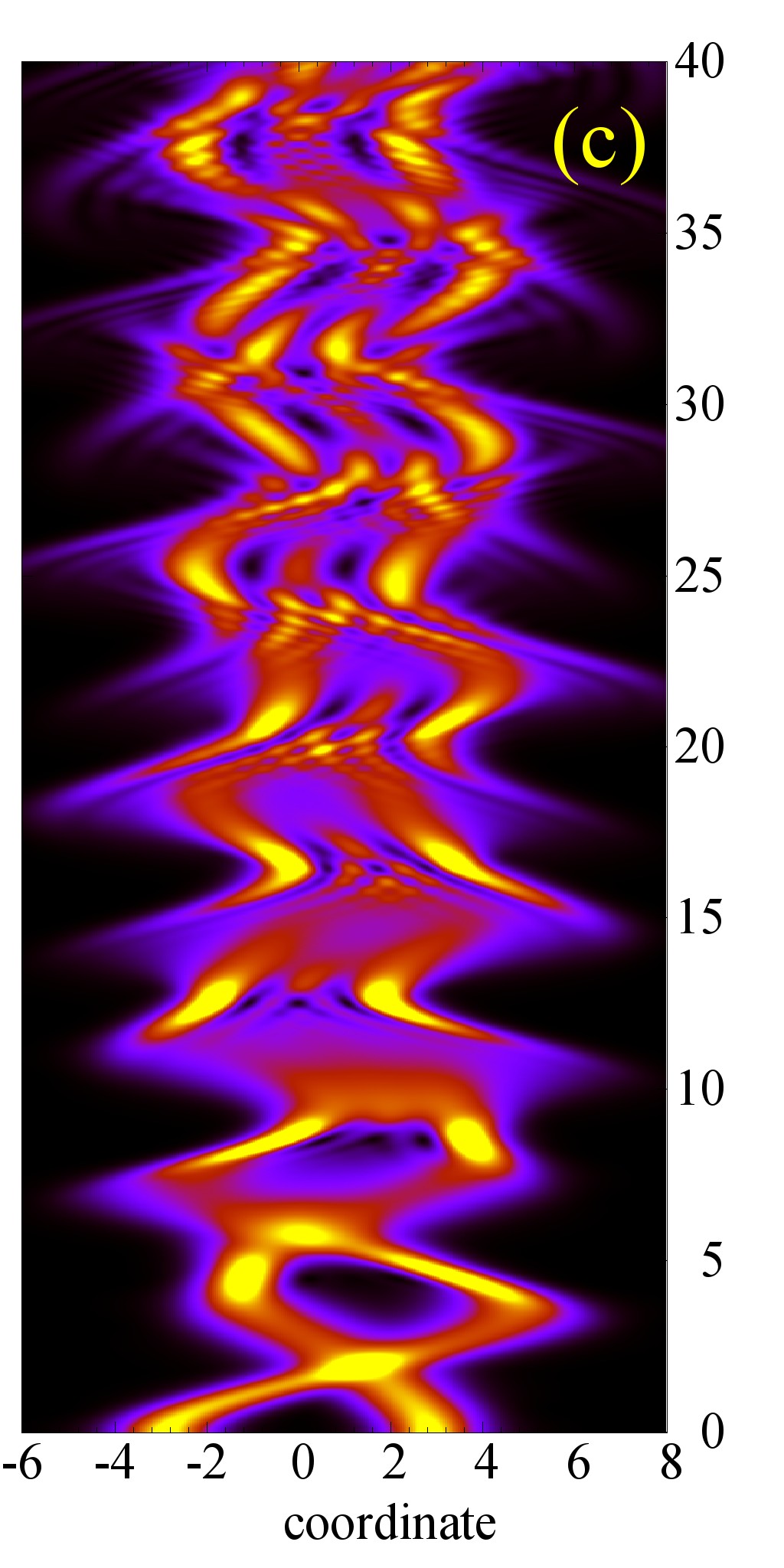}
\caption{(color online).
Two generic scenarios of many-body dynamics in 1D induced by a sudden displacement 
of the trap's origin $V(x)\!=\!x^2/2\!\to\!V(x\!-\!1)$ and a simultaneous quench of the inter-particle repulsion.
The initial state is the ground state of $N\!=\!100$ bosons confined in $V(x)$ with $\lambda_0\!=\!0.5$.
Minkovski-like space-time evolutions of the density are plotted for the scenarios activated by the displacement of the trap 
(a) without quench of the repulsion;
(b) with small increase of the repulsion $\lambda_0\!=\!0.5\!\to\!0.65$;
(c) with substantial decrease of the repulsion $\lambda_0\!=\!0.5\!\to\!0.1$.
Panels (a,b) reveal a first generic regime  --  non-violent, under-a-barrier dynamics.
Panel (c) represents a second, over-a-barrier regime with a highly-non-equilibrium, explosive changes of the density.
All quantities shown are dimensionless.}
    \label{fig1}
\end{figure}

Let us first take a two-hump two-fold fragmented system in 1D,
obtained as the ground state of $N\!=\!100$ bosons confined in $V(x)\!=\!x^2/2$ with $\lambda_0\!=\!0.5$, see Fig.~S1(b) of the supplemental material.
At $t\!=\!0$ we suddenly displace the trap with respect to its origin $V(x)\!\to\!V(x-1)$.
The computed density of the evolving many-body wave-packet
in a Minkovskii-like space-time representation is  depicted in Fig.~\ref{fig1}(a).
The main observation is that this manipulation of the trap induces only a non-violent many-body dynamics --
the two-hump topology and the two-fold fragmentation of the system persist
for all the presented times.

Now we enrich the dynamical scenario studied above by imposing onto the same initial system   
together with the sudden displacement of the trap at $t\!=\!0$ also a sudden small quench of the inter-particle repulsion form $\lambda_0\!=\!0.5\!\to\!0.65$.
The computed wave-packet evolution is depicted in a Minkovski-like manner in Fig.~\ref{fig1}(b).
The wave-packet dynamics reveals along with a relative simple motion induced by the trap displacement also new additional features.
The sub-clouds forming the wave-packet change their widths during the evolution back and fourth, i.e., they ``breath''. 
These breathings slightly disturb and modulate the perfect harmonic-like  oscillations of both sub-clouds.

Next, in a third scenario we take the same initial state as before,
suddenly displace the trap at $t\!=\!0$ but now significantly decrease the strength of the inter-particle repulsion form $\lambda_0\!=\!0.5\!\to\!0.1$.
At this value of the inter-particle interaction the ground state of the final system has a one-hump topology and
is fully condensed, see Fig.~S1(a) of the supplementary material.
So, this scenario can be considered as an attempt to make a super-fluid from an initially fragmented system.
The corresponding Minkovski-like evolution of the density is shown in Fig.~\ref{fig1}(c).
The wave-packet reveals highly-non-equilibrium explosive dynamics which is
accompanied by the formation of complicated oscillating patterns in the density.
The character of this dynamics differs drastically from the evolutions studied above, compare panels (a-b) with panel (c) of Fig.~\ref{fig1}.

The studied 1D many-body systems demonstrate two, generically different reactions to the applied manipulations with the external trapping and inter-particle interaction potentials -- a non-violent and a highly-non-equilibrium, explosive ones.
A main claim of this Letter is that these reactions are generic features of disturbed strongly interacting repulsive systems.
To confirm this generality we have extended the above reported 
manipulations with trap displacements and quenches of the inter-particle interactions to 2D and 3D setups, see Figs.~\ref{fig2},\ref{fig3} and the supplementary material for details.

A this point it is worthwhile to stress that all the multi-hump systems studied here are 
initially fragmented and remain fragmented during the non-violent and explosive time-evolutions.
In fragmented systems, there are no definite phase relations (correlation) between the sub-fragments, 
pretty much as in Mott-like states \cite{MI}. In fully condensed systems, in contrast, the phase between different humps is fixed.
Fig.~S2 of the supplemental material shows how one-body correlation functions \cite{RDM}
can be used to distinguish fragmented and condensed systems.

To understand the physics behind the non-violent and explosive regimes of the time-dependent many-body dynamics
of the trapped multi-hump multi-fold fragmented systems with strong finite-range inter-particle interactions,
we extend the concept of self-induced effective potentials
introduced in Ref.~\cite{Towards} to explain static properties of the multi-hump ground states.
As in the static case, we associate each hump with a well-isolated, now time-evolving sub-system $\phi_k(\br,t)$.
During the propagation, as one can see from the MCTDHB equations of motion, Eqs.~(\ref{MCTDHB_eq}),
each of these fragments ``feels'' the external trap potential and, also, 
effective time-dependent potential barriers $W_{jj}(\br,t)\!=\!\int |\phi_j(\br',t)|^2 W(\br\!-\!\br') d\r'$ induced by the other (counterpart) sub-clouds 
as a result of the inter-particle interaction.
The intensity, i.e., the height of the induced, time-dependent barriers, is proportional to the strength of the repulsion $\lambda_0$
and to the ratio  ${\{\bldrho(t)\}^{-1}_{jq}}\rho_{qksl}(t)$ of the involved elements of reduced two- and one-body density matrices, see MCTDHB equations, Eqs.~(\ref{MCTDHB_eq}).
In the case of an ideal fragmented state with well-localized fragments, this ratio is proportional to the number of particles residing in each fragment. For instance, in the case of a perfect two-fold fragmented system it approaches $N/2$, where $N$ is the total number of particles.

\onecolumngrid

\begin{figure}[h]
\includegraphics[width=5.75cm,angle=0]{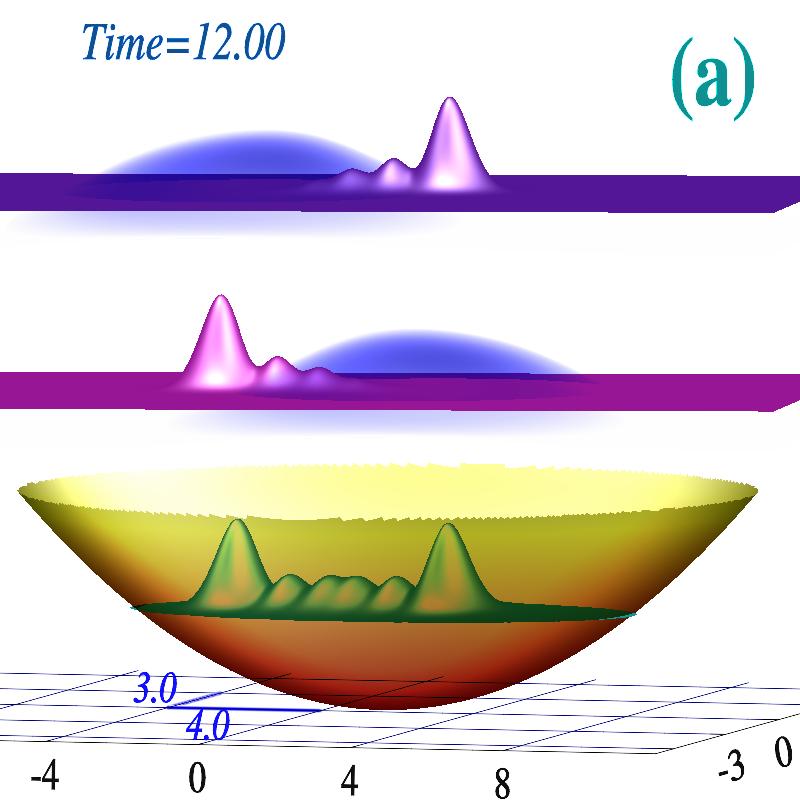}
\includegraphics[width=5.75cm,angle=0]{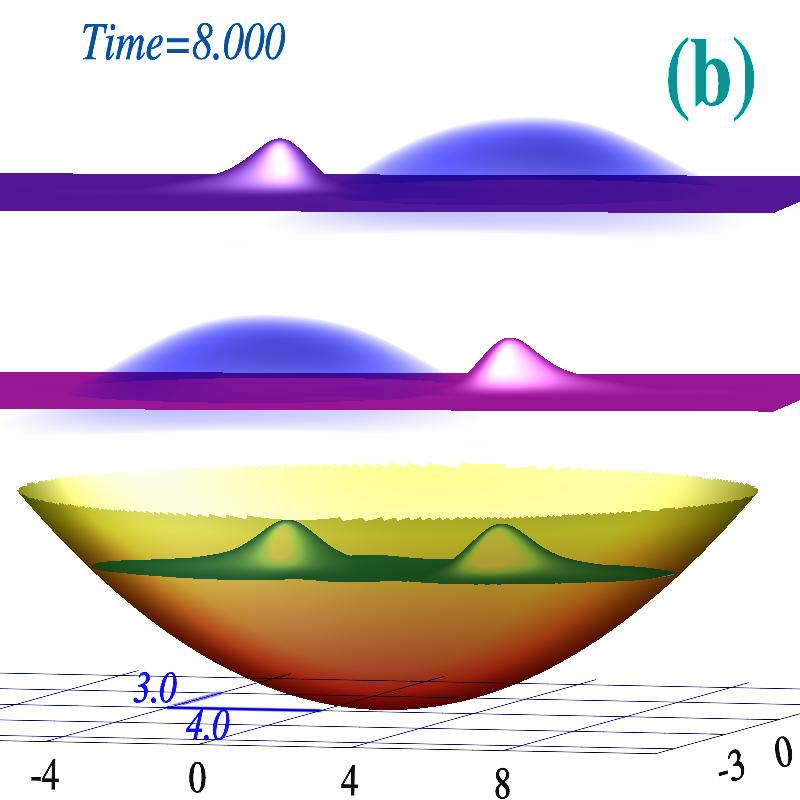}
\includegraphics[width=5.75cm,angle=0]{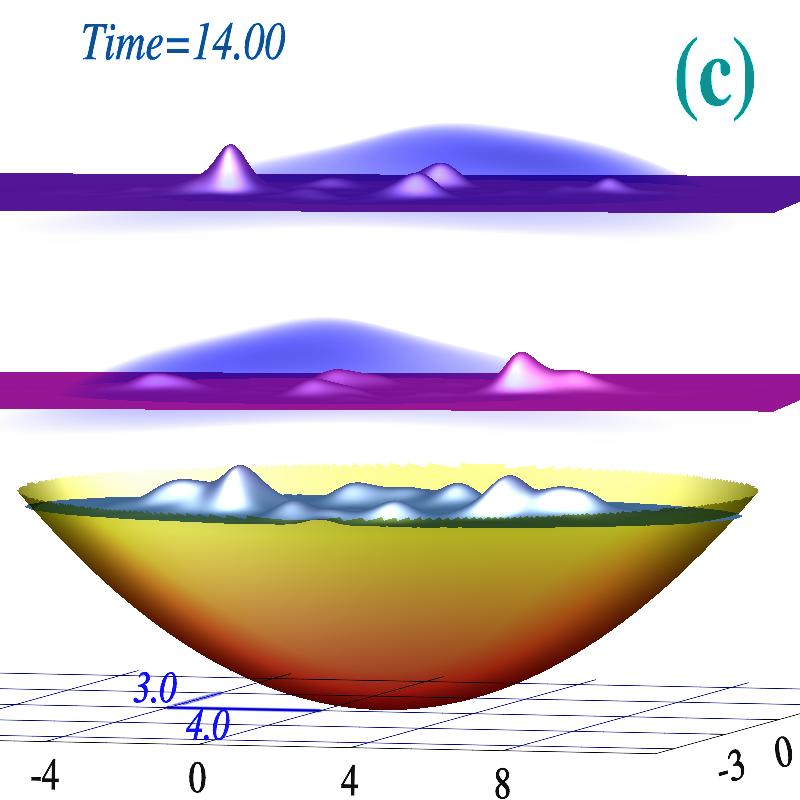}
\caption{(color online). Visualization of the concept of interaction-induced time-dependent 
barriers to explain two generic dynamical regimes; A 2D case.
Evolutions of a two-hump two-fold fragmented initial state induced by
a sudden displacement of the trap $V(x,y)\!=\!0.5x^2\!+\!1.5y^2\!\to\!V(x\!-\!1.5,y\!-\!0.5)$
with a simultaneous quench of the inter-particle repulsion:
(a) strong decrease $\lambda_0\!=\!0.5\!\to\!0.1$,  snap-shot at $t\!=\!12$;
(b) moderate increase $\lambda_0\!=\!0.5\!\to\!0.7$, snap-shot at $t\!=\!8$;
(c) stronger increase $\lambda_0\!=\!0.5\!\to\!0.8$, snap-shot at $t\!=\!14$.
Upper panels show the densities $|\phi_k(\r',t)|^2$ of the $k\!=\!L,R$  fragments [working MCTDHB orbitals, Eqs.~(\ref{MCTDHB_eq})] 
and the time-dependent barriers $\lambda_0N/2\!\int|\phi_j(\r',t)|^2W(\r\!-\!\r') d\r'$  induced by the complimentary $j\!=\!R,L$ sub-clouds.
Trapping potentials and computed density are shown on the lower panels.
The over-a-barrier dynamics (a,c) happens when the energy per particle of the out-of-equilibrium state is larger than 
the heights of the induced barriers,  otherwise the dynamics is under-a-barrier (b).
The induced barriers depicted in (a) have been multiplied by a factor of 40, for better visualization.
All quantities shown are dimensionless.}
    \label{fig2}
\end{figure}

\twocolumngrid

For illustrative purposes we examine and validate the applicability of this analysis in 2D setups where the dynamics is induced by a sudden displacement of the  trap potential  $V(x,y)\!=\!0.5x^2\!+\!1.5y^2\!\to\!V(x\!-\!1.5,y\!-\!0.5)$ 
with simultaneous quenches of the repulsion.
The initial state is the two-hump two-fold fragmented ground state, trapped in $V(x,y)$ with $\lambda_0\!=\!0.5$.
In the lower panels of Fig.~\ref{fig2} we depict snapshots of the evolving densities and respective trapping potential at several different time-slices 
for the above described scenarios of the trap displacement
and quenches of the repulsion:
from $\lambda_0\!=\!0.5\!\to\!0.1$, at $t=12$ in the Fig.~\ref{fig2}(a);
from $\lambda_0=0.5\!\to\!0.7$, at $t=8$ in the Fig.~\ref{fig2}(b); 
from $\lambda_0=0.5\!\to\!0.8$, at $t=14$ in the Fig.~\ref{fig2}(c).
The initial two-hump state studied here has a dominant ($99.9\%$) contribution from the $|N/2,N/2\rangle$ configuration,
indicating on an essentially perfect two-fold fragmentation of the system. During the propagation the fragmentation ratios change, of course,
but the systems still remain two-fold fragmented.
In the upper panels of Fig.~\ref{fig2} we plot snap-shots of the densities 
$|\phi_k(\r,t)|^2$ of the $k\!=\!L,R$ left/right sub-clouds  and
the effective time-dependent barriers $W_{jj}(\r,t)\!=\!\int |\phi_j(\r',t)|^2 W(\r\!-\!\r') d\r'$ induced by the counterpart $j\!=\!R,L$ (right/left) fragments at the same time-slices. To make the induced barriers visible in the case of a strong decrease of the interaction from $\lambda_0\!=\!0.5\!\to\!0.1$ depicted in the Fig.~\ref{fig2}(a),
we have scaled them by a factor of 40.
The full movies of the respective many-body dynamics are available in the supplemental  material \cite{SM}.

In Fig.~\ref{fig2} one can clearly distinguish two qualitatively different regimes of evolutions.
A non-violent one, plotted in the middle (b) sub-figure,
can be contrasted with highly-non-equilibrium ones, depicted in the left (a) and right (c) sub-figures.
To explore and confirm the existence of these two generic regimes observed in one- and two- dimensional setups
also in 3D, we plot in Fig.~\ref{fig3} an example of a highly non-equilibrium violent dynamics 
induced by a sudden displacement of the trap 
$V(x,y,z)\!=\!0.5x^2\!+\!1.5y^2\!+\!1.5z^2\!\to\!V(x\!-\!1.5,y\!-\!0.5,z\!-\!0.5)$ and strong decrease of the finite-range repulsion 
from $\lambda_0\!=\!0.5\!\to\!0.1$.
Here, to visualize the 3D functions we plot several isosurfaces of the density and an equipotential cut of the trap.
The snap-shots of the densities are taken at $t\!=\!0,2,3,6.7$. Full movies of this and two other 3D scenarios are provided in the supplemental material \cite{SM}.

Let us digest the physics of the above observed dynamical regimes.
In the non-violent evolutions, see e.g. Fig.~\ref{fig1}(a,b) for 1D and
Fig.~\ref{fig2}(b) for 2D, the
strong repulsion prevents an exchange of the particles between the sub-clouds, 
so they do not come to a proximity for a contact. 
The superposition of the induced time-dependent barriers and external trap
results in effective potentials which are high enough to confine the sub-clouds even in the case when they are moving.
Hence, the non-violent dynamics of the trapped multi-hump, multi-fold fragmented repulsive systems
appears when the manipulations exerted on the system are not strong enough to destroy or overcome 
the induced time-dependent barriers. 
So, the physics behind is an under-a-barrier dynamics.

Complimentary, the violent dynamics appears when the induced barriers are not high enough to keep the sub-clouds apart from each other.
Indeed, in the above studied sudden decreases of the inter-particle repulsion from $\lambda_0\!=\!0.5\!\to\!0.1$,
depicted in the Fig.~\ref{fig1}(c) for 1D,  Fig.~\ref{fig2}(a) for 2D, and Fig.~\ref{fig3} for 3D, 
reduce the heights of the induced barriers, which are proportional to $\lambda_0$.
The sub-clouds start to leak out and eventually 
become delocalized over the entire trap. 
The $\lambda_0\!=\!0.5\!\to\!0.8$ quench in 2D, depicted in the Fig.~\ref{fig2}(c),
formally leads to an increase of the heights of the induced barriers,
but, simultaneously, it pumps too much internal energy into the system.
At this new value of the inter-particle interaction strength the ground state is three-fold fragmented and, hence,
the shapes of the initial two-hump sub-clouds are no more optimal.
It also implies that the energy per-particle of each sub-cloud 
is larger than the heights of the induced time-dependent barriers. 
During the violent evolutions the available energy is redistributed between the excited states 
which have multi-node structures and can be delocalized over the entire trap.
As a result, the density reveals the observed highly violent explosive behavior.
So, the physics behind is a complex over-a-barrier dynamics.


\begin{figure}[b]
\includegraphics[width=3.5cm,angle=0]{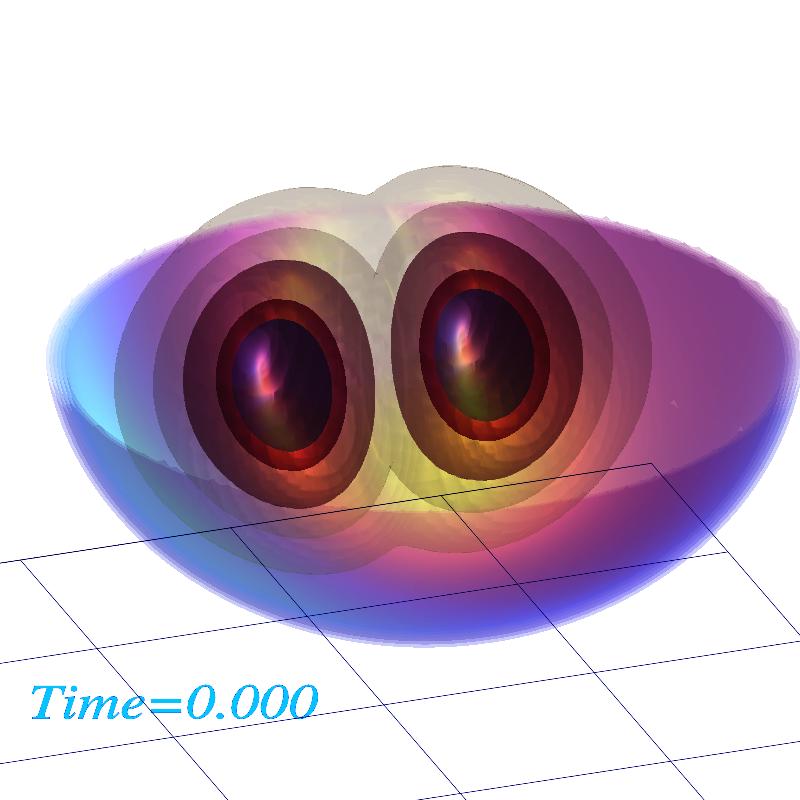}
\includegraphics[width=3.5cm,angle=0]{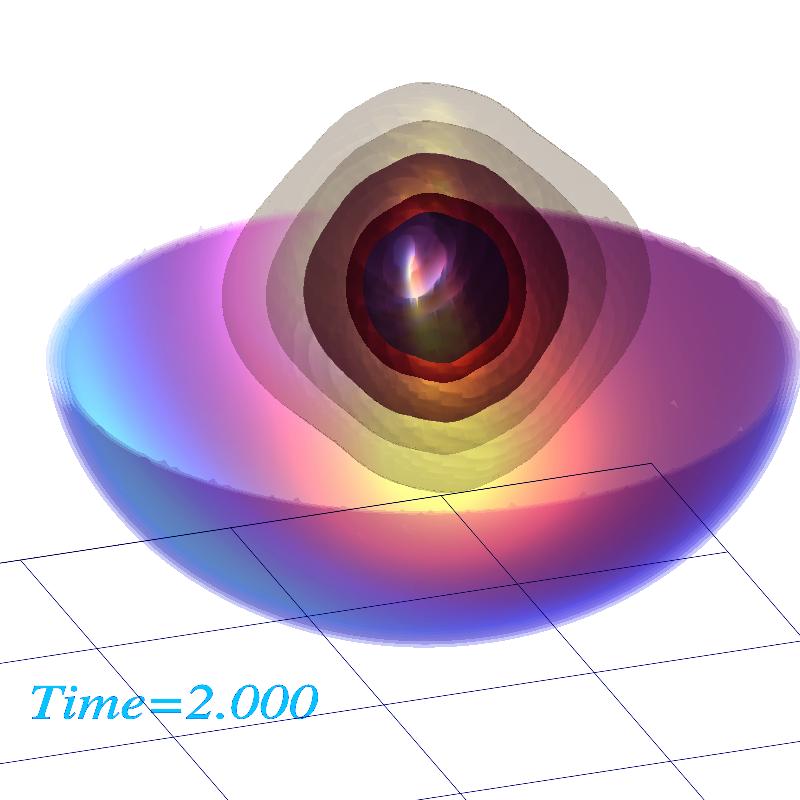}
\includegraphics[width=3.5cm,angle=0]{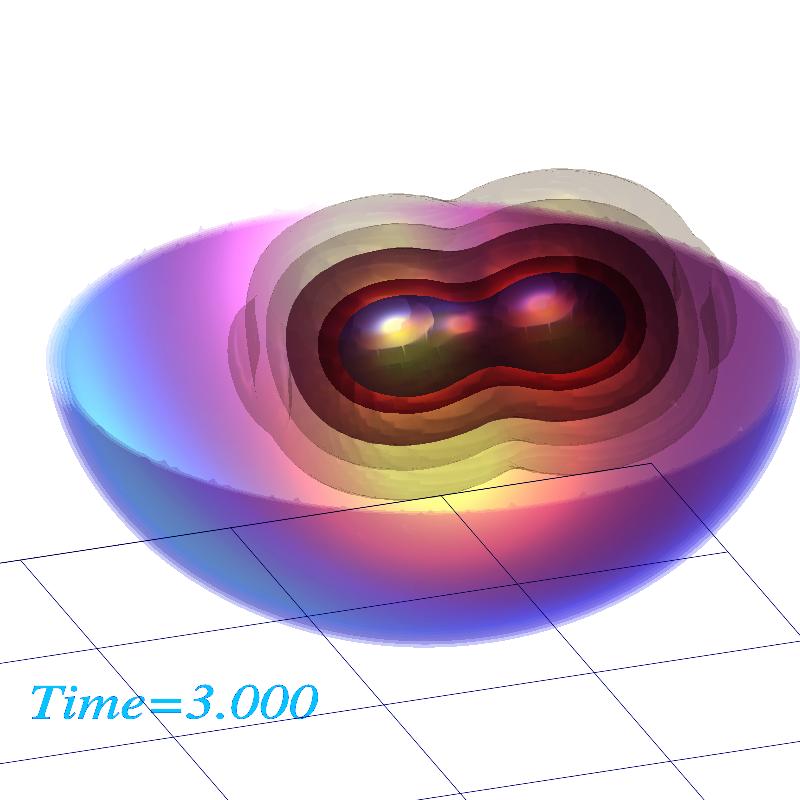}
\includegraphics[width=3.5cm,angle=0]{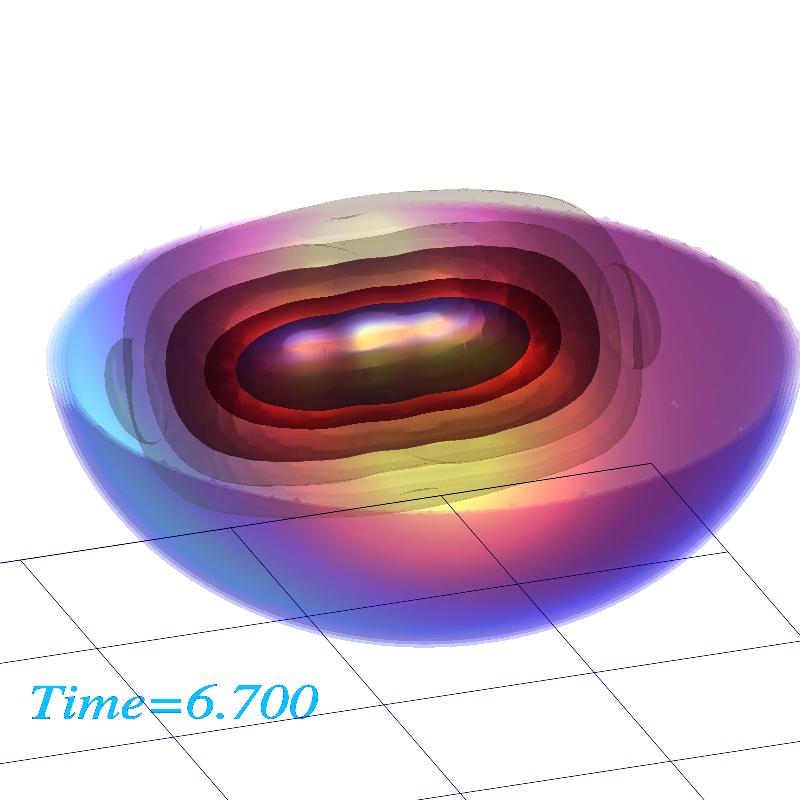}
\caption{(color online). Over-a-barrier regime in 3D. The evolution of a two-hump two-fold fragmented initial state
is induced by a sudden displacement of the trap 
$V(x,y,z)\!=\!0.5x^2\!+\!1.5y^2\!+\!1.5z^2 \to V(x\!-\!1.5,y\!-\!0.5,z\!-\!0.5)$
and strong, sudden decrease of the repulsion from $\lambda_0\!=\!0.5\!\to\!0.1$. 
To visualize the 3D functions we plot several isosurfaces of the density and an equipotential cut of the trap.
The snap-shots of the density are taken to show
initial ($t\!=\!0$), coalescing ($t\!=\!2$), penetrating ($t\!=3$) and highly-excited ($t\!=6.7$) momentary instances of the violent over-a-barrier dynamics.
All quantities shown are dimensionless.}
    \label{fig3}
\end{figure}

Now, we are able to deduce a set of practical recommendations on possible experimental 
preparations and manipulations of the multi-hump, multi-fold fragmented states.
(i) Once prepared, these states remain
very stable and robust with respect to possible imperfectness of experimental setups.
(ii) A protocol where the non-interacting system is first prepared and then the interaction is diabatically quenched seems 
to be ineffective because, it would lead to explosive dynamics.
(iii) Finally, a formation of the systems with a desired number of humps (fragments) can be provoked and controlled
by pre-imposing a weak optical lattice of the required periodicity from the beginning of the quench process.
By switching it off afterwords,  one could, in principle, induce only a non-destructive non-violent under-a-barrier dynamics.

Concluding, we have shown that the physics behind the two 
hitherto found
generic regimes of the many-body non-equilibrium dynamics 
of  trapped ultracold Bose systems with strong repulsive finite/long range interactions 
is driven by the mechanism of interaction-induced time-dependent barriers.
The non-violent dynamics of the first, under-a-barrier regime
is contrasted to a highly-non-equilibrium explosive quantum many-body dynamics 
of the second,  over-a-barrier regime.
The generality of the discovered time-dependent physics is verified in one-, two-, and three-spatial dimensions.

%


Computation time on the bwGRiD, HLRS and K100 clusters are greatly acknowledged. A partial financial support by the DFG is acknowledged.
%

%

\newpage

\onecolumngrid

\def\beq{\begin{equation}}
\def\eeq{\end{equation}}
\def\beqn{\begin{eqnarray}}
\def\eeqn{\end{eqnarray}}
\def\souligne#1{$\underline{\smash{ \hbox{#1}}}$}

\def\r {{\bf r}}
\def\n {{\bf n}}
\def\d {{\bf d}}

\def\C {{\bf C}}
\def\U {{\bf U}}
\def\D {{\bf D}}
\def\T {{\bf T}}
\def\O {{\bf O}}

\def\r {\vec{\bf r}}
\def\r {{\bf r}}
\def\br {{\bf r}}
\def\bR {{\bf R}}
\def\bcalH {\mbox{\boldmath $\mathcal H$}}
\def\ul {\underline}
\def\bcalH {\mbox{\boldmath $\mathcal H$}}

\def\brho {\mbox{\boldmath $\rho$}}

\def\bphi {\mbox{\boldmath $\phi$}}


\title{Supplemental Material\\ $ \ $ \\ 
Generic regimes of quantum many-body dynamics \\ of trapped bosonic systems with strong repulsive interactions}


%
%
%
%
%
%

\newpage
\thispagestyle{empty}

\addtocounter{figure}{-3}

\renewcommand{\figurename}{Figure S\hglue -0.12 truecm}

\makeatletter
\renewcommand*{\@biblabel}[1]{[S#1]}
\makeatother

\section{Supplemental Material}

In this supplementary material we first show that
the inter-particle interaction of the $W(\bR)=1/{((|\br-\br'|/D)^{n}+1})$ shape, used throughout the present study,
appears naturally in the context of the ``Rydberg-dressed'' atoms.
So, it is possible to verify all the predicted physics within presently available experimental setups. 
Next, we demonstrate how to control the number of humps and fragmentation ratio in the ground state 
of the bosonic system with such an inter-particle interaction trapped in simple barrier-less traps, also see Ref.~[10].
We also show how to distinguish fragmented systems studied here from the condensed ones, by means of the correlation functions.
Finally, we provide the full movies of the time-dependent evolutions in 2D and 3D setups, discussed in the main text.

\section*{On possible experimental realization of the predicted phenomena}

Let us start from the issue of possible experimental realizations of the predicted physics.
To observe the predicted effects the first necessary condition is that the length/range of the inter-particle interaction
should be comparable with the length of trapping potentials. 
The second necessary condition is that the interaction should be strong enough.

We expect to observe the formation of few-hump fragmented ground states 
in trapped ultra-cold systems made of ``Rydberg-dressed'' atoms [5-7]. 
Let us discuss how to realize the predicted multi-hump muti-fold fragmented physics in ``Rydberg-dressed" systems.
The off-resonant optical coupling of ground state atoms to highly excited Rydberg states [6, 7] 
allows one to modify and control the shape and strength of effective two-body interactions.
The advantage of this technique is that a small component of the Rydberg state is admixed to the ground state of atoms,
providing thereby an additional degree of manipulation of the interaction strength.

The shape of the potential energy of the two many-electron atoms excited to a Rydberg state depends very much on details of the electronic structure.
The most common long-range behavior, however, is of a standard Van der Waals ($C_6/R^6$) type with possible admixture of terms of other degrees, 
e.g., a pure dipole interaction  ($C_3/R^3$). These additional contributions are responsible for the tails of the interaction potentials.
The ``dressing" of two Rydberg atoms with Van der Waals type of interaction results, see
Refs.~[6, 7],  
in the two-body effective inter-particle potential:
$$
\tilde{W}_{dd}(|\tilde{r}_i-\tilde{r}_j|)=\frac{\tilde{C}_6}{\tilde{r}_{ij}^6+R_c^6}=\frac{\Omega^4}{8 \Delta^3} \frac{\hbar}{(\tilde{r}_{ij}/R_c)^6+1}.
 \eqno{\mathrm {(S1)}}
$$
Let us make a change of variables $\tilde{r}_{i} = r_i l_{\omega}$, where $\omega$ is the frequency of the external trap.
In the units of the external confining potential's length, $l_{\omega}=\sqrt{\frac{\hbar}{m \omega}}$,
the inter-particle interaction reads
$$\lambda_0W(|r_i-r_j|)=\frac{\lambda_0}{(r_{ij}/D)^6+1}. \eqno{\mathrm {(S2)}}$$ 
Where $\lambda_0=\frac{\hbar\Omega^4}{(2 \Delta)^3}\frac{1}{\hbar \omega}$ is the rescaled interaction strength
and $D=R_c$. 
In the present study, we have used one-, two- and three-dimensional versions of inter-particle interactions
of a similar shape $\lambda_0/(\frac{|\br-\br'|}{D})^{n}+1)$ of half-width $D=4$ with $n=4$.
This degree of the inter-particle repulsion function corresponds to an ``intermidiate'' situation between the pure dipole-dipole and Van der Waals interactions. 
We have used it to demonstrate the generality of the observed physics,
which holds for $n=3$ and $n=6$ as well.

The strength $\lambda_0$ of the inter-particle interaction depends on 
experimentally tunable parameters: $\Omega$  -- a two-photon Rabi frequency between the involved atomic levels,
$\Delta$ -- detunings of lasers with respect to the atomic transitions, and the external confinement $\omega$.
The ``screening" constant $D\equiv R_c=\frac{C_6}{2 \hbar |\Delta|}$
defines the critical distance, below which the interaction is constant, i.e., originates to the blockade phenomenon [5].
This range depends on the detuning $\Delta$ and on 
the pure spectroscopic $C_6$ coefficient governed by the electronic structure of the excited state, i.e., it
can be manipulated by a proper choice of an atomic Rydberg level.

What are the presently available/reachable experimental conditions for Rydberg excitations? 
In recent experiment, see Ref.~[8],  with $^{87}$Rb Bose-Einstein condensates,
a few tens of Rydberg atoms in a quasi-1D trap have been successfully detected.
The reported blockade radii were between 5-\SI{15}{\micro\metre} 
while the radial dipole trap frequencies used were around \SI{100}{\hertz}, resulting in a radial length of
order of 1-\SI{2}{\micro\metre}, i.e., a ratio $l_{\omega} \sim 0.1 R^{Rb}_{c}$, where $R^{Rb}_{c}$ is  the ``screening" constant of the Rb atoms.
It means that the range of the interaction was larger than the the size of the trap in the transverse direction. 
The strength of the interaction of the dressed Rydberg atoms
is defined by the $\frac{\Omega^4}{8 \Delta^3}$ ratio and can be tuned between 0.1-\SI{10}{\kilo\hertz}, implying that 
in a \SI{100}{\hertz} trap the corresponding dimensionless interaction strength $\lambda_0=\frac{\hbar\Omega^4}{(2 \Delta)^3}\frac{1}{\hbar \omega}$
can be $\sim$ 1-100. 
So, formally, the lengths/ranges and interaction strength needed to observe the predicted multi-hump muti-fold fragmented states and their dynamics 
are already reachable with ultra-cold ``dressed" Rydberg atoms.

\section*{Versatility of the control on the ground state}

Let us now demonstrate that ground state properties of the trapped bosonic system 
with a finite-range repulsive interaction can be controlled 
by varying either the strength of the inter-particle repulsion $\lambda_0$, the tightness/strength 
of harmonic confinement $\omega$,  or by changing the number $N$ of the trapped particles. For more details see Ref.~[10].
Here, we explicitly consider 1D bosons with finite-range inter-particle interaction function $\lambda_0/((|x-x'|/D)^{4}+1)$ with $D=4$.
The ground state of $N=100$ bosons trapped in a standard 
$V(x)=0.5 \omega^2x^2$ harmonic trap, i.e., with $\omega=1.0$, 
at weak repulsion $\lambda_0=0.1$ is super-fluid, i.e., condensed and has the one-hump density, see Fig.~S1(a). 
By increasing the strength of the repulsion one gets at $\lambda_0=0.5$ a two-fold fragmented ground state 
with two-hump density, see Fig.~S1(b). At even stronger repulsion, $\lambda_0=0.8$,
the ground state is three-fold fragmented with a three-hump density profile, see Fig.~S1(c).
Alternatively, three-fold fragmented states with a three-hump density profiles can be prepared in a weaker external trap $\omega=0.75$ for $N=100$ with $\lambda=0.5$, see Fig.~S1(f)
or in a harmonic trap with $\omega=1.0$, but for $N=300$ and $\lambda_0=0.3$, see Fig.~S1(d).
Finally, Fig.~S1(e) shows that the system of $N=100$ with $\lambda=0.5$ placed in an asymmetric trap 
$V(x)\!\!=\!\!\left\{0.5x^2:x\!<\!0; 0.01x^4\!:\!x\!\ge\!0\right\}$,
results in a two-fold fragmented ground state with an asymmetric two-hump density,
verifying thereby that the symmetry of the trap does not play a significant role here.

\begin{figure}[]
\includegraphics[width=10.0cm,angle=0]{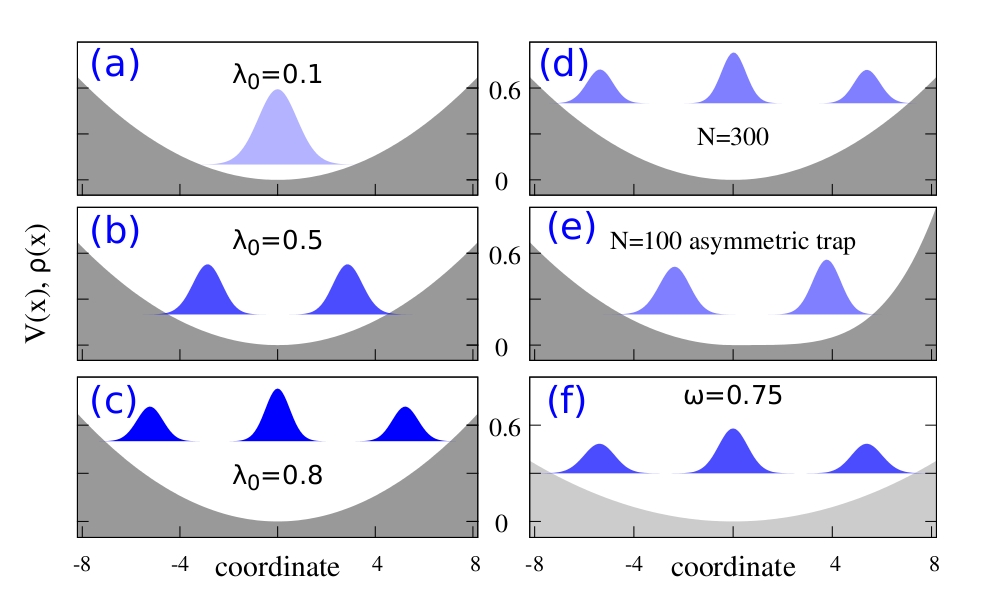}
\caption{(color online). Control on the multi-hump structure can be gained by varying either 
strength of the inter-particle interaction $\lambda_0$, the tightness/strength 
of external confinement $\omega$,  or by changing the total number $N$ of the trapped particles.
The ground state of the system with $N=100$ interacting via a finite-range $\lambda_0/((|x-x'|/D)^{4}+1)$ with $D=4$ 
trapped in a standard $V(x)=0.5 \omega^2x^2$ harmonic trap, i.e., with $\omega=1.0$,  at weak repulsion $\lambda_0=0.1$
is super-fluid  and has the one-hump density (a). At stronger interactions 
it becomes multi-fold fragmented with two-hump densities at $\lambda_0=0.5$  (b)  and  three-hump at $\lambda_0=0.8$ (c).
Alternatively,  the three-hump densities can be obtained 
by preparing the system of $N=300$  with $\lambda_0=0.3$ 
in the standard harmonic trap (d), or
the system of $N=100$ with $\lambda=0.5$ in a weaker external trap with $\omega=0.75$ (f).
The system of $N=100$ with $\lambda=0.5$ placed in an asymmetric trap $V(x)\!\!=\!\!\left\{0.5x^2:x\!<\!0; 0.01x^4\!:\!x\!\ge\!0\right\}$
results in an asymmetric density of the fragmented state (e).
All quantities shown are dimensionless.}
    \label{fig1}
\end{figure}

The above demonstrated versatility of control functions implies that in experiments a desired fragmentation scenario
can be equivalently obtained (i) by squeezing/opening the external trap; 
(ii) by changing the number of trapped bosons; or (iii) by varying the strength of the repulsion. 

The time-dependent studies reported in the main text 
have equipped us with detailed time-dependent understanding of the above mentioned control functions.
Namely, the topology of a once prepared system cannot be easily and smoothly changed by 
a simple, sudden quench of the repulsion.
Indeed, we have shown that attempts to bring an initially two-fold fragmented system to the super-fluid, one-hump regime 
via strong decrease of the interaction,
or to a three-fold fragmented state via strong increase of the interaction
have resulted in explosive dynamics. 
The systems with different number of humps have different topologies
and live in different configurational (Fock) spaces, implying that to bring one system to another one 
we have to ``transfer'' a macroscopic number of the particles from one fragment (hump) to the other fragments which is, clearly, a very non-trivial many-body task. Furthermore, it is easier to realize the systems with a few humps  and many particles per hump than
the many-hump system with a few particles per hump. 
Since the height of the induced barrier is proportional to the interaction strength and
to the number of particles residing in the fragment, 
the humps with a larger number of particles are more stable
with respect to exchange and loss of particles, i.e., to the temperature.

\section*{Correlation functions}

Now we would like to demonstrate that all the multi-hump systems studied here are initially fragmented
and remain fragmented during the non-violent and explosive time-evolutions.
In Fig.~S2 we depict the one-body correlation functions 
$|g^{(1)}(x',x;t)|^2$ computed
for the 1D dynamical scenarios of Fig.~1(a,b,c) at time-slice of $t\!=\!13$.
These correlation functions have typical characteristic patterns of two-fold fragmented structures, see [18]. 
The bright squares in Fig.~S2(a) mean that, within each sub-cloud, the bosons are coherent (condensed), while the
dark off-diagonal squares imply that there are no definite phase relations (correlation) between the sub-fragments,
pretty much as in Mott-like states [19].
In fully condensed systems, in contrast, the phase between different humps is fixed, 
i.e., the correlation function of the evolving fully condensed system would always be flat and bright everywhere, i.e., $|g^{(1)}(x',x;t)|^2\!=\!1$.

\begin{figure}[h]
\includegraphics[width=8.5cm,angle=0]{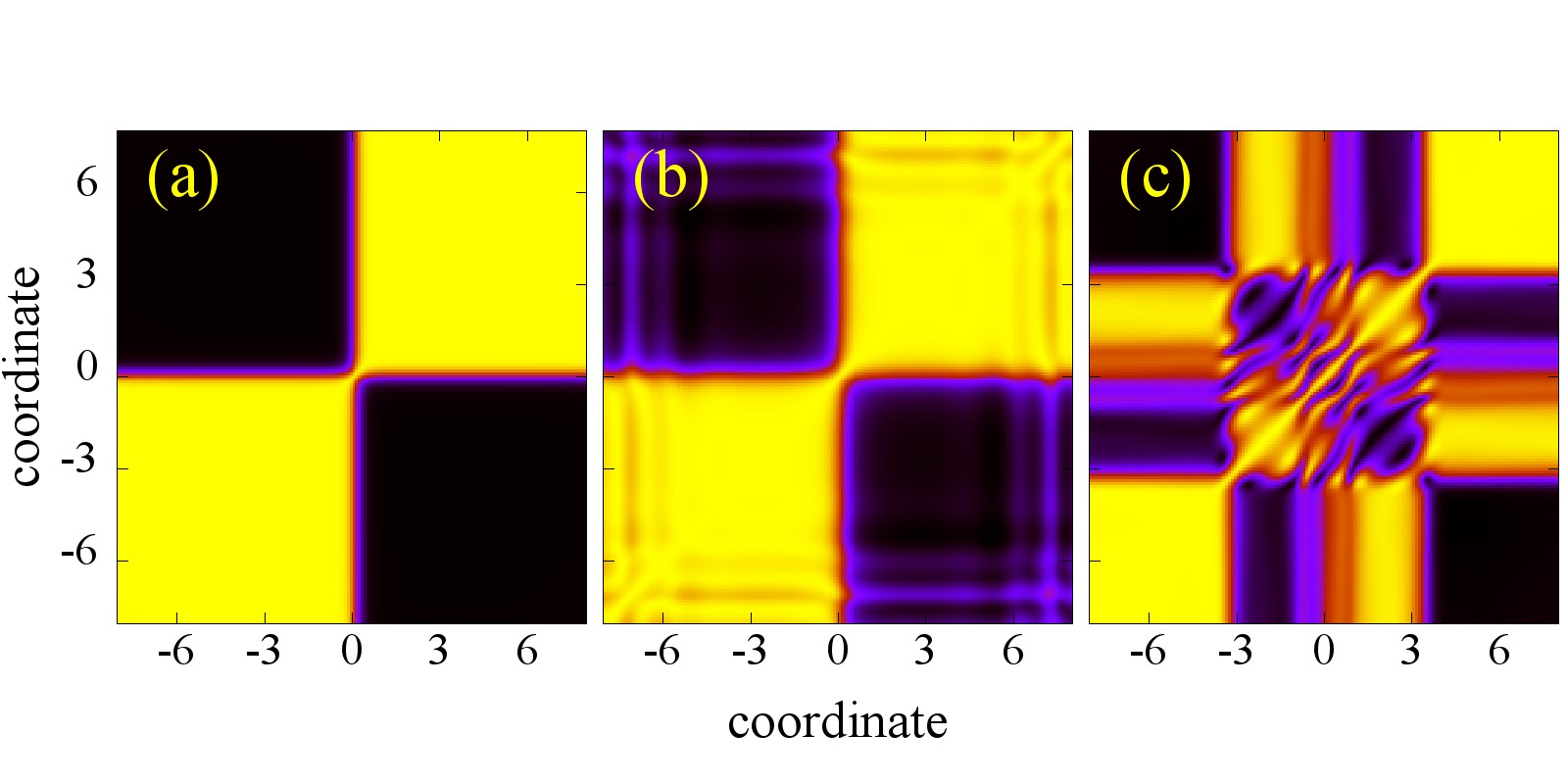}
\caption{(color online).
One-body correlation functions $|g^{(1)}(x',x;t)|^2$  taken at the same time-slice 
$t=13$ for the 1D dynamical scenarios  depicted in Fig.~1(a,b,c) of the main text.  
The two-fold fragmented structures of the evolving wave-packets 
survive during the non-violent (a,b) and explosive (c) dynamics.
All quantities shown are dimensionless.}
    \label{fig2}
\end{figure}

\section*{List of the movies with dynamics of 2D and 3D systems}

\begin{description}
\item[2D] \verb|2D_MCTDHB_quench_to_l_0d1.mpg|  file
contains the movie exemplifying the over-a-barrier evolution of a two-hump two-fold fragmented initial state in a 2D setup induced by
a sudden displacement of the trap $V(x,y)\!=\!0.5x^2\!+\!1.5y^2\!\to\!V(x\!-\!1.5,y\!-\!0.5)$
with a simultaneous strong reduction of the inter-particle repulsion $\lambda_0\!=\!0.5\!\to\!0.1$.
One slide of this movie, corresponding to a snap-shot at $t\!=\!12$, is depicted in Fig.~3(a).

\item[2D] \verb|2D_MCTDHB_quench_to_l_0d7.mpg| file 
contains the movie exemplifying the under-a-barrier evolution of the above scenario but
with a moderate increase of the inter-particle repulsion $\lambda_0\!=\!0.5\!\to\!0.7$.
One slide of this movie, corresponding to a snap-shot at $t\!=\!8$, is depicted  in Fig.~3(b).

\item[2D] \verb|2D_MCTDHB_quench_to_l_0d8.mpg| file 
contains the movie exemplifying the over-a-barrier evolution of the above scenario but
with a stronger increase of the inter-particle repulsion $\lambda_0\!=\!0.5\!\to\!0.8$.
One slide of this movie, corresponding to a snap-shot at $t\!=\!14$, is depicted in Fig.~3(c).

\item[3D] \verb|3D_MCTDHB_quench_to_l_0d1.mpg|  file
contains the movie exemplifying the over-a-barrier evolution of a two-hump two-fold fragmented initial state in a 3D setup induced by
a sudden displacement of the trap $V(x,y,z)\!=\!0.5x^2\!+\!1.5y^2\!+\!1.5z^2 \to V(x\!-\!1.5,y\!-\!0.5,z\!-\!0.5)$
and strong sudden decrease of the repulsion from $\lambda_0\!=\!0.5\!\to\!0.1$. 
In Fig.~4 four slides of this movie, corresponding to snap-shots at $t\!=\!0,2,3,6.7$, are depicted.

\item[3D] \verb|3D_MCTDHB_quench_to_l_0d7.mpg|  file
contains the movie exemplifying the under-a-barrier evolution of the above 3D scenario but
with a moderate increase of the inter-particle repulsion $\lambda_0\!=\!0.5\!\to\!0.7$.

\item[3D] \verb|3D_MCTDHB_quench_to_l_0d8.mpg|  file
contains the movie exemplifying over-a-barrier evolution of the above 3D scenario but
with a stronger increase of the inter-particle repulsion $\lambda_0\!=\!0.5\!\to\!0.8$.

\end{description}

\newpage
\thispagestyle{empty}

\addtocounter{figure}{0}

\renewcommand{\figurename}{Figure S\hglue -0.12 truecm}

%
%
%
%


\end{document}